\documentclass[twocolumn,superscriptaddress,floatfix,showpacs,aps]{revtex4}
\usepackage{amsmath,amssymb,epsfig,color,xcolor}
\pdfoutput=1
\usepackage{graphicx}
\usepackage{dcolumn}
\usepackage{bm}
\usepackage{color}
\usepackage{multirow}

\begin{document}

\title{Raman spectroscopy of Na$_3$Co$_2$SbO$_6$}

\author{Yu.S. Ponosov}
\affiliation{Institute of Metal Physics, S. Kovalevskaya Street 18, 620108 Ekaterinburg, Russia}

\author{ E.V. Komleva}
\affiliation{Institute of Metal Physics, S. Kovalevskaya Street 18, 620108 Ekaterinburg, Russia}
\affiliation{Ural Federal University, Mira St. 19, 620002 Ekaterinburg, Russia}

\author{E.A. Pankrushina}
\affiliation{A.N. Zavaritsky Institute of Geology and Geochemistry, Ural Branch of the Russian Academy of Sciences, Ak. Vonsovskogo St. 15, 620016, Ekaterinburg, Russia}
\affiliation{Ural Federal University, Mira St. 19, 620002 Ekaterinburg, Russia}

\author{D. Mikhailova}
\affiliation{Institute for Complex Materials, Leibniz Institute for Solid State and Materials Research (IFW) Dresden, 01069 Dresden, Germany}

\author{S.V. Streltsov}
\affiliation{Institute of Metal Physics, S. Kovalevskaya Street 18, 620108 Ekaterinburg, Russia}
\affiliation{Ural Federal University, Mira St. 19, 620002 Ekaterinburg, Russia}

\begin{abstract}
Raman spectroscopy together with density functional calculations were used to study lattice dynamics in a layered honeycomb cobaltite Na$_3$Co$_2$SbO$_6$, which can host a field-induced phase related with the Kitaev physics. We show that there develops an additional mode well above Neel temperature (at $\approx 200$K) at 525 cm$^{-1}$, which origin can be related to an electronic excitation to one of $j_{3/2}$ doublets. Moreover, our theoretical calculations demonstrate that the highest frequency intensive mode related to the oxygen vibrations is very sensitive to type of the magnetic order. Thus, we propose to use the softening of this mode as a hallmark of the transition to a fully polarized regime, which is stabilized in Kitaev materials in strong magnetic fields.
\end{abstract}

\maketitle

\date{\today}

\section{Introduction}
Competition of a bond-directional exchange interaction in some  transition metal compounds leads to a strong frustration. As a result, even at the lowest temperatures a system has no long-range magnetic order and has a finite entropy. Such an interaction naturally appears in the Kitaev model described by the Hamiltonian 
\begin{eqnarray}
H = \sum_{\langle ij \rangle_{\gamma} }  K^{\gamma} \hat S^{\gamma}_i \hat S^{\gamma}_j,
\end{eqnarray}
where different spin components $\gamma = \{ x,y,z\}$ for three symmetry inequivalent nearest neighbours (numerated by site indexes $i$ and $j$) on the honeycomb lattice turn out to be coupled by the exchange constants $K^{\gamma}$~\cite{Kitaev}. This is an extremely rare situation in the spin physics, when not only the ground state of a Hamiltonian, but also the excitation spectrum can be calculated exactly.

Materials, which can be a physical realization of this model, have become one of the central topics in the condensed matter physics for the last decade \cite{Takagi2019, Khomskii2021, Trebst2022}. Indeed, dynamical structure factor \cite{Knolle2014} and spin response in Raman spectra \cite{Knolle2014-2, Nasu2016} have been calculated and, moreover, decent anomalies in these observables have been detected. While there are still strong debates on their origin, the Raman spectroscopy turned out to be a very efficient method of studying Kitaev materials. In particular, a wide continuum experimentally observed in Raman spectra can arguably serve a hallmark of Kitaev physics.

In this paper we present results of Raman measurements for one of the honeycomb cobaltites Na$_3$Co$_2$SbO$_6$ considered as possible candidates for Kitaev physics. Experimental results are supplemented by the density functional theory (DFT) calculations. Theoretical consideration demonstrates that phonon modes turn out to be rather sensitive to the magnetic ordering and this can be used in further experiments. 

To go further on let us first  recapitulate basic physical properties of  Na$_3$Co$_2$SbO$_6$. Co is 2+ in this material with electronic configuration $3d^7$ and spin $S=3/2$. Co ions form a layered honeycomb structure and their magnetic moments order at $T_N \sim 5-8$ K (depending on the sample quality). The magnetic ground state corresponds to a so-called antiferromagnetic (AFM) zigzag structure, when ferromagnetic (FM) spins are ordered in zigzag fashion~\cite{Yan2019}. Effective magnetic moment is $5.2-5.5 \mu_B$~\cite{Viciu2007,Wong2016,Yan2019}, suggesting substantial contribution of the orbital moment. In spite of the fact that Na$_3$Co$_2$SbO$_6$ orders magnetically it was recently shown that a moderate magnetic field of 1-2~T suppresses the long-range magnetic order and leads to a field-induced state, which can retain some features of Kitaev physics~\cite{Vavilova2023}. First Raman data only for 160 - 260  cm$^{-1}$ were demonstrated in supplemental materials of Ref. \cite{Li2022}, while in this paper we present spectra in a wide frequency range.

\section{Experiment and calculation details.}
We used pollycrystalline samples described in Ref.~\cite{Vavilova2023}. Raman measurements in the temperature range of 80 to 300 K were performed in backscattering geometry from the policrystalline sample  Na$_3$Co$_2$SbO$_6$ using an RM1000 Renishaw microspectrometer equipped with a 532 nm solid-state laser and 633 helium–neon laser. Respective Linkam stage was used for temperature variation. Very low power (up to 1 mW) was used to avoid local heating of the sample. A pair of notch filters with a cut-off at 60 cm$^{-1}$ were used to reject light from the 633 nm laser line. To reach as close to the zero frequency as possible, we used a set of three volume Bragg gratings (VBG) at 532 nm excitation to analyze the scattered light. The resolution of our Raman spectrometer was estimated to be 2–3 cm$^{-1}$. To obtain information about the symmetry of the observed excitations, polarization measurements were tried in two geometries, with parallel   and  mutually perpendicular polarizations  of  the incident and scattered light.

All calculations have been performed using VASP\cite{Kresse4}. We utilized generalized gradient approximation (GGA) as proposed by Perdew, Burke, and Ernzerhof (PBE) ~\cite{Perdew} and 6$\times$6$\times$6 mesh of Brillouin zone in the reciprocal space. Crystal structure was taken from \cite{Yan2019}. Cutoff energy was chosen to be 500 eV. Convergence criteria for electronic and ionic cycles were set up to $10^{-7}$ and $10^{-6}$ eV, respectively.

We used DFT+U approach, which takes into account strong Coulomb correlations in a mean-field way~\cite{Liechtenstein1995}. Hubbard $U$ and Hund's $J_H$ were chosen to be $U=7$ eV, $J_H=1$ eV, similar what was used previously for other cobalitites~\cite{Zvereva2016,Maksimov2022}. 
Phonon frequencies at $\Gamma$-point were computed using the density functional perturbation theory (DFPT)~\cite{Gados2006,Baroni2001}. Note, that simulation of zigzag AFM implies doubling of the unit cell in the DFT calculations, which results in ``unphysical'' doubling of phonon modes.

\section{Experimental results.}
Na$_3$Co$_2$SbO$_6$ crystallizes in a monoclinic structure (space group C2/m) which implies optical phonons in the  center of the Brillouin zone are of the following symmetry:
\begin{eqnarray}
7A_g + 7A_u+8B_g + 11B_u.
\end{eqnarray}
These are 7A$_g$ and 8B$_g$ phonons, which can be observed by Raman spectroscopy.

\begin{figure}[t!]
\centering
\includegraphics[width=1\columnwidth]{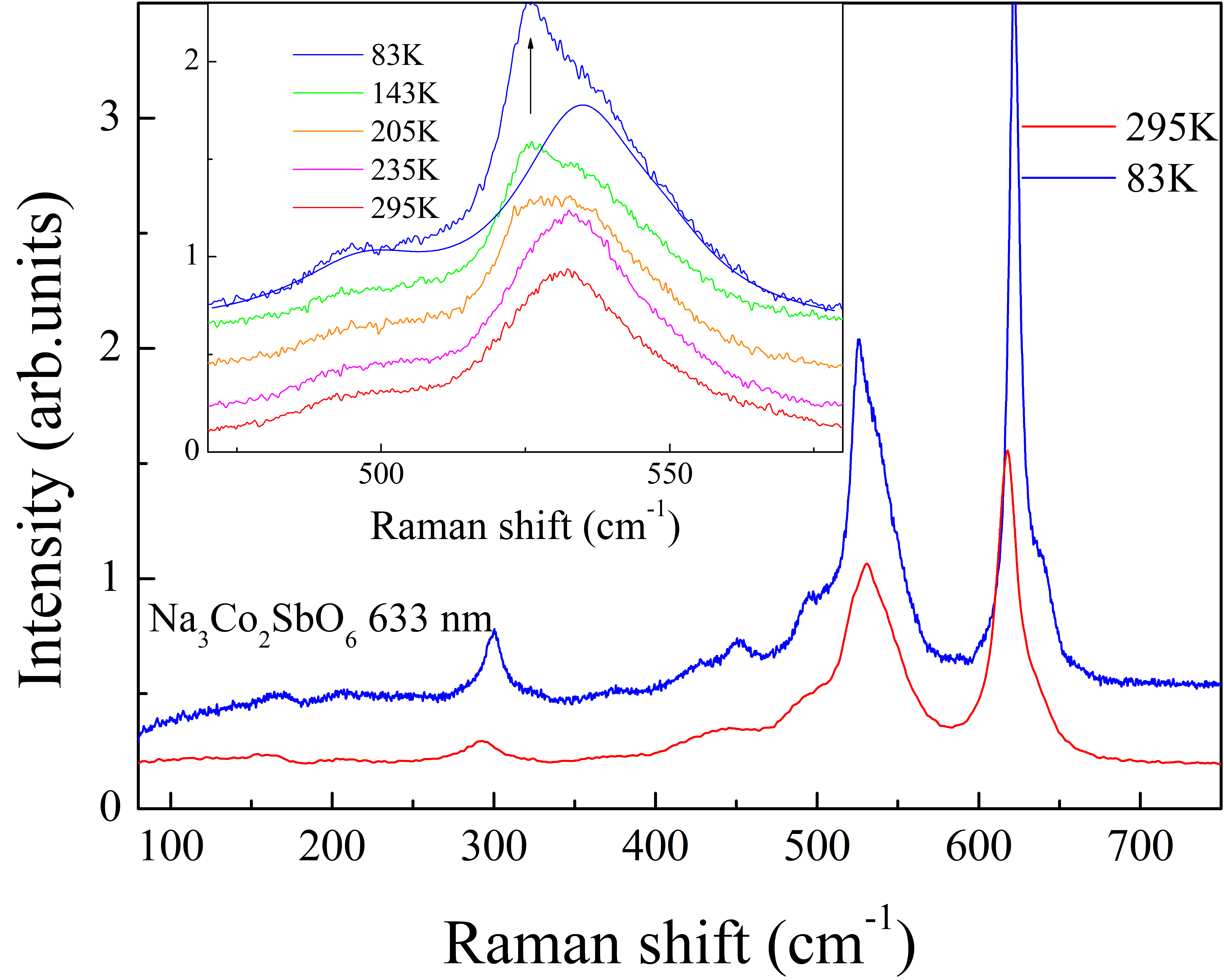}
\caption{Raman spectra of  Na$_3$Co$_2$SbO$_6$ measured at 83K and 295K. Insert shows temperature evolution of the spectral range where line near 525 cm$^{-1}$ appears.}
\label{Raman-experiment}
\end{figure}

The measured room temperature spectra show 11 lines at frequencies 121, 159, 207, 292, 371, 444, 495, 528, 543, 618, and 636 cm$^{-1}$. The widest lines (up to 60 cm$^{-1}$) are several times larger than the narrowest ones. This can be due to some disorder in the powder sample. This is also evidenced by temperature measurements (Fig.~\ref{Raman-experiment}). As the temperature decreases, most of the lines narrow only slightly; the accompanying hardening suggests the influence of anharmonic contributions.  The background on which the phonon lines are superimposed is structureless in the region up to 5000 cm$^{-1}$, but its intensity changes when measured at different points of the sample in accordance with the intensity of the phonon spectrum. The exact shape of phonon lines is difficult to determine due to their large width and the overlap of some lines with others.

An interesting effect was observed in the region of 200 K, where a new narrow mode appeared at 525 cm$^{-1}$ with decreasing temperature (insert in Fig.~\ref{Raman-experiment}). Since there is no information about structural transitions at this temperature in Na$_3$Co$_2$SbO$_6$, the origin of this effect is unclear. One of the possible reasons can be excitations from the ground state doublet $j=1/2$ to one of $j=3/2$ doublets. Indeed, our preliminary exact diagonalization calculations for isolated Co$^{2+}$ ion with parameters extracted from DFT (Hamiltonian was obtained by Wannier function projection, spin-orbit coupling constant was taken to be 60 meV, $U=7$ eV, and $J_H=1$ eV) give excitation energies 15 and 60 meV.

\begin{figure}[t!]
\centering
\includegraphics[width=1\columnwidth]{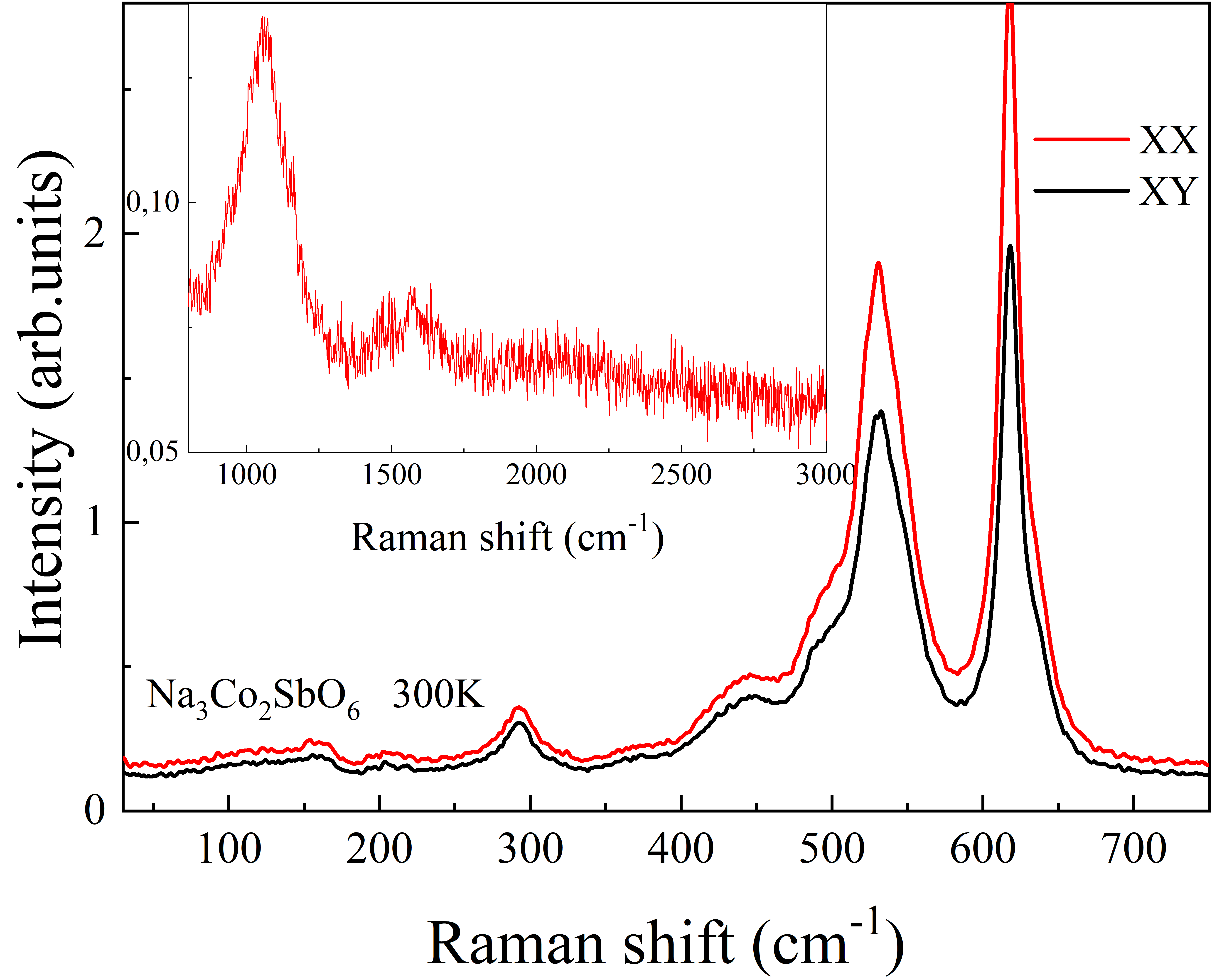}
\caption {Raman spectra of  Na$_3$Co$_2$SbO$_6$ measured in XX and XY scattering geometries at 300K. High-frequency spectral range containing two-phonon excitation is shown in insert.}
\label{Raman-experiment-HF}
\end{figure}

The spectra turned out to be identical upon excitation by both laser lines at 532 and 633 nm. An attempt to evaluate the line depolarization in the samples showed the minimal effect for the highest frequency lines at 528 and 618 cm$^{-1}$, which is typical for lines of A$_g$ symmetry, see Fig.~\ref{Raman-experiment-HF}. For a more accurate experimental identification of the symmetry, phonon lineshapes, background and extra line origin, measurements on a single crystal and at lower temperatures are required. In the high-frequency region, the spectra show broad bands at 1060, 1570 and possibly 2080 cm$^{-1}$ (insert in Fig.~\ref{Raman-experiment-HF}), the origin of which is presumably associated with phonon repetitions (high-order scattering from phonons in the region of 530 cm$^{-1}$). At the same time, it is not clear why there is no repetition of the second intensive phonon mode with frequency 618 cm$^{-1}$.

\section{Theory.}
\begin{figure}[t!]
\centering
\includegraphics[width=1\columnwidth]{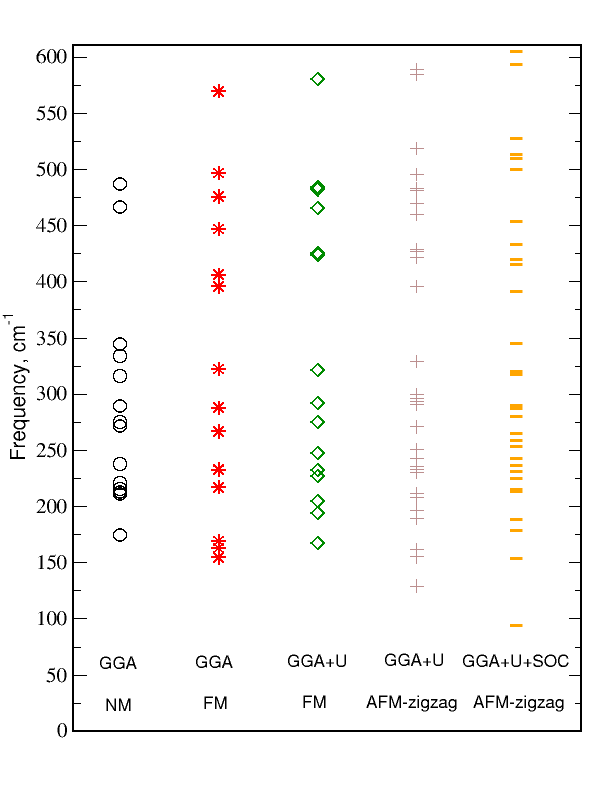}
\caption{Frequencies of phonon modes in $\Gamma-$point as obtained by different DFT calculations and extracted from the experiment (EXP). NM stands for non-magnetic, FM - ferromagnetic, AFM- antiferromagnetic, SOC - spin-orbit coupling. \label{DFT-freq}}
\end{figure}

Results of theoretical simulations performed within DFT calculations together with experimental frequencies are summarized in Fig.~\ref{DFT-freq}. All modes are of the $A_g$ or $B_g$ symmetry. First of all, one can see that phonon frequencies obtained in the non-spin-polarized calculation (non-magnetic case) differ strongly from the experimental data in both the low and high frequency regions. Taking spin degrees of freedom into account leads to a pronounced effect on phonon frequencies: even accounting for a ferromagnetism significantly improves agreement with experiment, where the highest phonon modes are observed at $\sim$600 cm$^{-1}$. Moreover, the calculation for experimentally observed zigzag-AFM shifts both the highest and lowest frequencies is even closer to the experimental Raman line. This highlights the importance of the spin-lattice coupling in Na$_3$Co$_2$SbO$_6$. It has to be mentioned at this point that the zigzag type of AFM requires doubling of the unit cell and this results in increase of number of frequencies (this effect is not physical; it can be avoided by a back-folding procedure, which requires however calculation of full phonon spectrum).

Clear dependence of highest- and lowest-frequency modes on type of the magnetic structure can be used by the Raman spectroscopy to study magnetic field effects. Indeed hardening of this mode signals on transition to a fully polarized state. It would be very interesting to investigate such an effect and frequency dependence of these modes in an intermediate field range, where zigzag AFM would have been suppressed and a field-induced phase with possible Kitaev physics starting to emerge. 

Another point to be mentioned is a modification of the phonon spectrum by the spin-orbit coupling. This interaction further improves agreement for the highest-frequency mode: AFM-zigzag GGA+U gives 588.8 cm$^{-1}$, while AFM-zigzag GGA+SOC+U -- 600.0 cm$^{-1}$. It also modifies low-lying modes, but this can not be compared with experiment, since not all modes are seen in the measurements.

\begin{figure}[h!]
\centering
\includegraphics[width=1\columnwidth]{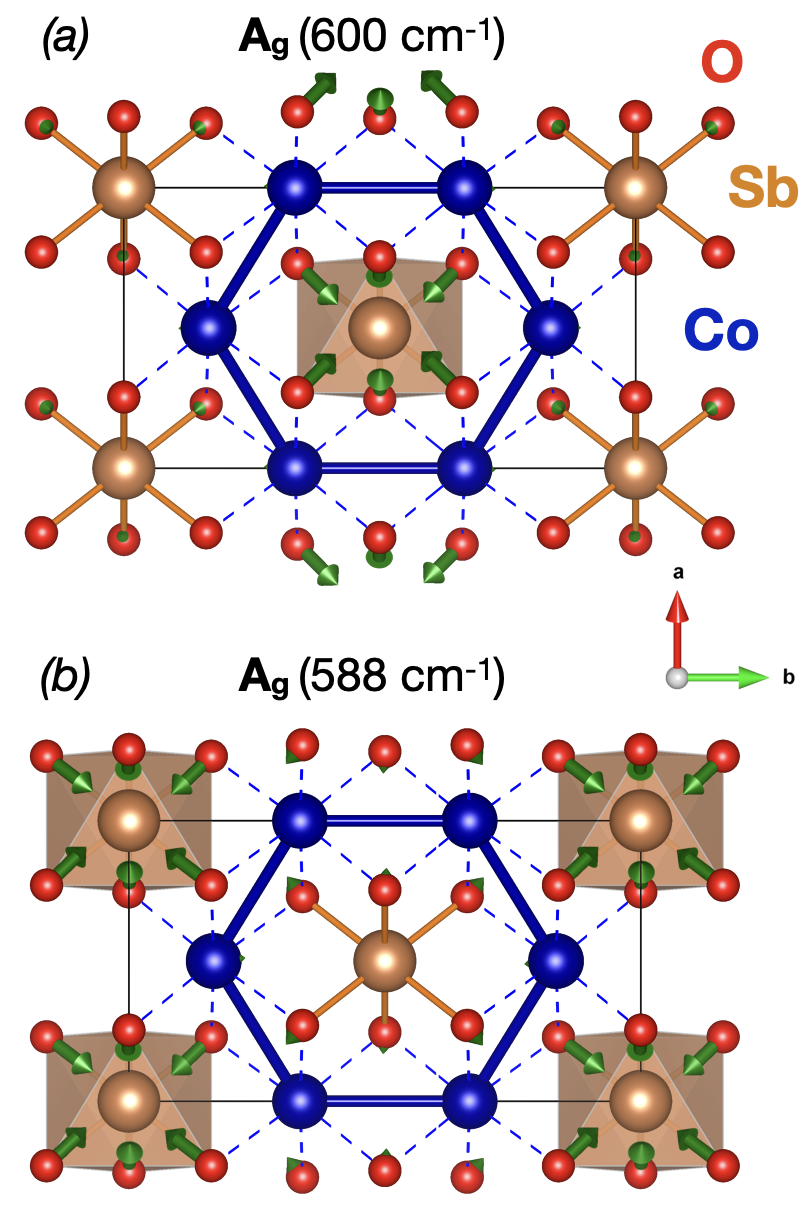}
\caption{Atomic displacement patterns for two highest $A_g$ modes ((a) for 600~cm$^{-1}$ and (b) for 588~cm$^{-1}$) calculated in GGA+U+SOC for the zigzag magnetic structure  shown in the $ab$ plane. The orange octahedra demonstrate those with the largest vibration amplitude of oxygen atoms. Other SbO$_6$ clusters shown with lines also have the same type of vibrations but with the significantly smaller amplitude. As one can see, this $A_g$ mode corresponds to the compression of the oxygen octahedra around the Sb atoms. \label{Distortions}}
\end{figure}

It is interesting to get into the origin of hardening of the highest phonon modes in AFM configuration. Fig.~\ref{Distortions} demonstrates displacements for these two modes. One can see that they correspond to stretching distortions of SbO$_6$ octahedra, which are in the centers of Co hexagons. In case of AFM zigzag order magnetostriction leads to decrease of Sb-Co distances (from 3.13$\times$4 \AA~and 3.11$\times$2 \AA~to 3.07$\times$4 \AA~and 3.08$\times$2 \AA). This explains frequency growth of the stretching mode. 

\section{Conclusions.}
 Lattice dynamics in the layered honeycomb cobaltite Na$_3$Co$_2$SbO$_6$ considered as a possible candidate to the Kitaev physics realization was investigated by Raman spectroscopy and DFT calculations. Raman spectra do not reveal formation of high-energy continuum observed in some Kitaev materials \cite{Knolle2014-2,Nasu2016}. This can be due the fact that our measurements were performed at rather high temperatures and without a magnetic field, which suppresses a long-range magnetic order moving the system towards Kitaev phase. Nevertheless, we observe formation of an additional peak at $\sim$ 525 cm$^{-1}$ below 200K, which could be associated with development of a short-range magnetic correlations. The extra peaks above 1000 cm$^{-1}$ typical for some other honeycomb lattice layered compounds and associated with high-order scattering of phonons are found. 
 
 Interestingly, phonon frequencies appeared to strongly depend on the magnetic arrangement of Co ions. Our results demonstrate the importance of spin-lattice coupling in this kind of materials and we argue that this effect can be used to distinguish various magnetic phases observed in future experiments. In addition, the spin-orbit coupling turns out to be important not only for correct description of magnetic properties, but it also significantly improves theoretical description of lattice dynamics. 

\section{Acknowledgements.} We are extremely grateful to T. Vasilchikova for help with samples and to P. Maksimov and E. Vavilova for various stimulating discussions on Na$_3$Co$_2$SbO$_6$. The work was supported by the Russian Science Foundation via project RSF 23-12-00159.

\end{document}